\documentclass[10pt,journal,compsoc]{IEEEtran}

\usepackage{epsfig}
\usepackage{graphicx}
\usepackage{subfigure}
\usepackage{amsmath}
\usepackage{amsfonts}
\usepackage{multirow} 
\usepackage{mathrsfs}

\ifCLASSOPTIONcompsoc
\else
\fi

\ifCLASSINFOpdf
\else
\fi

\begin{document}
%
\title{Recent Advance in Content-based Image Retrieval: A Literature Survey}
%
%
%

\author{Wengang Zhou,~
        Houqiang Li,~
        and Qi Tian ~\IEEEmembership{Fellow,~IEEE}

\IEEEcompsocitemizethanks{
\IEEEcompsocthanksitem Wengang Zhou and Houqiang Li are with the CAS Key Laboratory of Technology in Geo-spatial Information Processing and Application System, Department of Electronic Engineering and Information Science, University of Science and Technology of China, Hefei, 230027, China. \protect\\ E-mail: \{zhwg, lihq\}@ustc.edu.cn.

\IEEEcompsocthanksitem Qi Tian is with the Department of Computer Science, University of Texas at San Antonio, San Antonio, TX, 78249, USA. \protect\\ E-mail: qitian@cs.utsa.edu.}
}

\IEEEtitleabstractindextext{
\begin{abstract}
The explosive increase and ubiquitous accessibility of visual data on the Web have led to the prosperity of research activity in image search or retrieval. With the ignorance of visual content as a ranking clue, methods with text search techniques for visual retrieval may suffer inconsistency between the text words and visual content. Content-based image retrieval (CBIR), which makes use of the representation of visual content to identify relevant images, has attracted sustained attention in recent two decades. Such a problem is challenging due to the intention gap and the semantic gap problems. Numerous techniques have been developed for content-based image retrieval in the last decade. The purpose of this paper is to categorize and evaluate those algorithms proposed during the period of 2003 to 2016. We conclude with several promising directions for future research.
\end{abstract}

}

\maketitle

\IEEEdisplaynontitleabstractindextext

%
\IEEEpeerreviewmaketitle

\section{Introduction}
With the universal popularity of digital devices embedded with cameras and the fast development of Internet technology, billions of people are projected to the Web sharing and browsing photos. The ubiquitous access to both digital photos and the Internet sheds bright light on many emerging applications based on image search. Image search aims to retrieve relevant visual documents to a textual or visual query efficiently from a large-scale visual corpus. Although image search has been extensively explored since the early 1990s~\cite{rui1998relevance}, it still attracts lots of attention from the multimedia and computer vision communities in the past decade, thanks to the attention on scalability challenge and emergence of new techniques. Traditional image search engines usually index multimedia visual data based on the surrounding meta data information around images on the Web, such as titles and tags. Since textual information may be inconsistent with the visual content, content-based image retrieval (CBIR) is preferred and has been witnessed to make great advance in recent years.

In content-based visual retrieval, there are two fundamental challenges, \emph{i.e.,} \emph{intention gap} and \emph{semantic gap}. The intention gap refers to the difficulty that a user suffers to precisely express the expected visual content by a query at hand, such as an example image or a sketch map. The semantic gap originates from the difficulty in describing high-level semantic concept with low-level visual feature~\cite{alzu2015semantic}~\cite{li2016socializing}~\cite{lin2015semantics}. To narrow those gaps, extensive efforts have been made from both the academia and industry.

From the early 1990s to the early 2000s, there have been extensive study on content-based image search. The progress in those years has been comprehensively discussed in existing survey papers~\cite{smeulders2000content}~\cite{lew2006content}~\cite{liu2007survey}. Around the early 2000s, the introduction of some new insights and methods triggers another research trend in CBIR. Specially, two pioneering works have paved the way to the significant advance in content-based visual retrieval on large-scale multimedia database. The first one is the introduction of invariant local visual feature SIFT~\cite{Lowe:IJCV04}. SIFT is demonstrated with excellent descriptive and discriminative power to capture visual content in a variety of literature. It can well capture the invariance to rotation and scaling transformation and is robust to illumination change. The second work is the introduction of the Bag-of-Visual-Words (BoW) model~\cite{sivic2003video}. Leveraged from information retrieval, the BoW model makes a compact representation of images based on the quantization of the contained local features and is readily adapted to the classic inverted file indexing structure for scalable image retrieval.

Based on the above pioneering works, the last decade has witnessed the emergence of numerous work on multimedia content-based image retrieval ~\cite{nister2006scalable}~\cite{philbin2007object}~\cite{jegou2008hamming}~\cite{zhou2011large}~\cite{sivic2003video}~\cite{chum2007total}
~\cite{philbin2008lost}~\cite{chum2008near}~\cite{wu2009bundling}~\cite{zhou2010spatial}~\cite{chum2011total}~\cite{zhang2011GVP}~\cite{zhang2012qsrank}
~\cite{he2012mobile}~\cite{arandjelovic2012three}~\cite{zhang2012query}~\cite{tian2011building}~\cite{zhou2010BiSpace}~\cite{zhang2009descriptive}
~\cite{zhang2010building}~\cite{zhou2011latent}. Meanwhile, in industry, some commercial engines on content-based image search have been launched with different focuses, such as Tineye\footnote{http://tineye.com/}, Ditto\footnote{http://ditto.us.com/}, Snap Fashion\footnote{https://www.snapfashion.co.uk/}, ViSenze\footnote{https://www.visenze.com}, Cortica\footnote{http://www.cortica.com/}, \emph{etc}.  Tineye is launched as a billion-scale reverse image search engine in May, 2008. Until January of 2017, the indexed image database size in Tineye has reached up to 17 billion. Different from Tineye, Ditto is specially focused on brand images in the wild. It provides an access to uncover the brands inside the shared photos on the public social media web sites.

Technically speaking, there are three key issues in content-based image retrieval: image representation, image organization, and image similarity measurement. Existing algorithms can also be categorized based on their contributions to those three key items.

Image representation originates from the fact that the intrinsic problem in content-based visual retrieval is image comparison. For convenience of comparison, an image is transformed to some kind of feature space. The motivation is to achieve an implicit alignment so as to eliminate the impact of background and potential transformations or changes while keeping the intrinsic visual content distinguishable. In fact, how to represent an image is a fundamental problem in computer vision for image understanding. There is a saying that ``An image is worth a thousand words''. However, it is nontrivial to identify those ``words''. Usually, images are represented as one or multiple visual features. The representation is expected to be descriptive and discriminative so as to distinguish similar and dissimilar images. More importantly, it is also expected to be invariant to various transformations, such as translation, rotation, resizing, illumination change, \emph{etc}.

In multimedia retrieval, the visual database is usually very large. It is a nontrivial issue to organize the large scale database to efficiently identify the relevant results of a given query. Inspired by the success of information retrieval, many existing content-based visual retrieval algorithms and systems leverage the classic inverted file structure to index large scale visual database for scalable retrieval. Meanwhile, some hashing based techniques are also proposed for indexing in a similar perspective. To achieve this goal, visual codebook learning and feature quantization on high-dimensional visual features are involved, with spatial context embedded to further enrich the discriminative capability of the visual representation.

Ideally, the similarity between images should reflect the relevance in semantics, which, however, is difficult due to the intrinsic ``semantic gap'' problem. Conventionally, the image similarity in content-based retrieval is formulated based on the visual feature matching results with some weighing schemes. Alternatively, the image similarity formulations in existing algorithms can also be viewed as different match kernels~\cite{Tolias2013to}.

In this paper, we focus on the overview over research works in the past decade after 2003. For discussion before and around 2003, we refer readers to previous survey~\cite{smeulders2000content}~\cite{lew2006content}\cite{liu2007survey}. Recently, there have been some surveys related to CBIR~\cite{zhang2013image}~\cite{alzu2015semantic}~\cite{li2016socializing}. In~\cite{zhang2013image}, Zhang \emph{et al.} surveyed image search in the past 20 years from the perspective of database scaling from thousands to billions. In~\cite{li2016socializing}, Li~\emph{et al.} made a review of the state-of-the-art CBIR techniques in the context of social image tagging, with focus on three closed linked problems, including image tag assignment, refinement, and tag-based image retrieval. Another recent related survey is referred in~\cite{alzu2015semantic}. In this work, we approach the recent advance in CBIR with different insights and emphasize more on the progress in methodology of a generic framework.

In the following sections, we first briefly review the generic pipeline of content-based image search. Then, we discuss five key modules of the pipeline, respectively. After that, we introduce the ground-truth datasets popularly exploited and the evaluation metrics. Finally, we discuss future potential directions and conclude this survey.

\section{General flowchart overview}

\begin{figure*}
\centering
\includegraphics[width = 15.0cm]{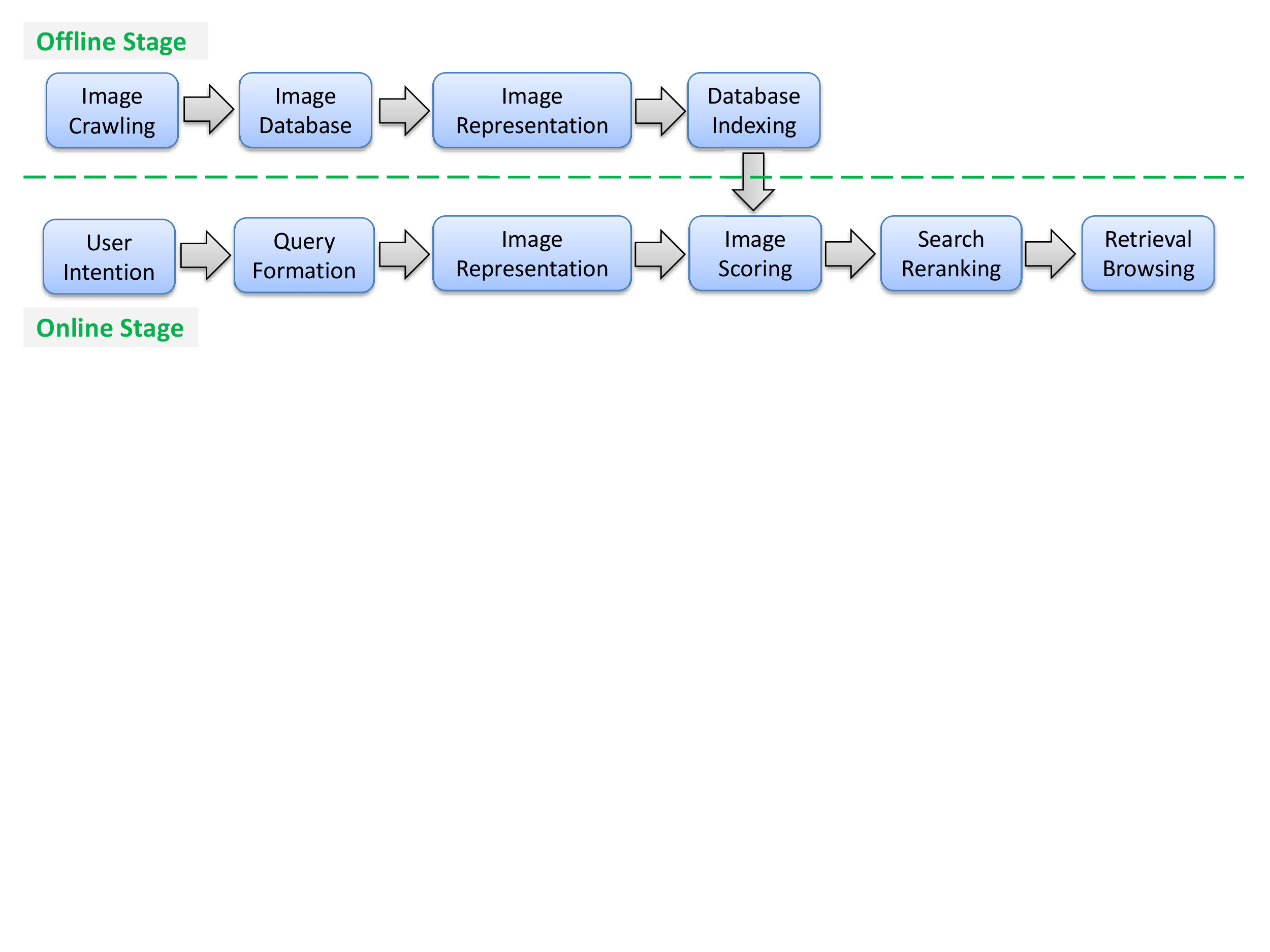}
\caption{The general framework of content-based image retrieval. The modules above and below the green dashed line are in the off-line stage and on-line stage, respectively. In this paper, we focus the discussion on five components, \emph{i.e.}, query formation, image representation, database indexing, image scoring, and search reranking.}
\label{fig:pipeline}
\end{figure*}

Content-based image search or retrieval has been a core problem in the multimedia field for over two decades. The general flowchart is illustrated in Fig.~\ref{fig:pipeline}. Such a visual search framework consists of an off-line stage and an on-line stage. In the off-line stage, the database is built by image crawling and each database image is represented into some vectors and then indexed. In the on-line stage, several modules are involved, including user intention analysis, query formation, image representation, image scoring, search reranking, and retrieval browsing. The image representation module is shared in both the off-line and on-line stages. This paper will not cover image crawling, user intention analysis~\cite{tang2012intentsearch}, and retrieval browsing~\cite{moghaddam2004visualization}, of which the survey can be referred in previous work~\cite{lew2006content}~\cite{datta2008image}. In the following, we will focus on the other five modules, \emph{i.e.}, query formation, image representation, database indexing, image scoring, and search reranking.


In the following sections, we make a review of related work in each module, discuss and evaluate a variety of strategies to address the key issues in the corresponding modules.

\section{Query Formation}
\begin{figure*}
  \centering
  \includegraphics[width=12 cm]{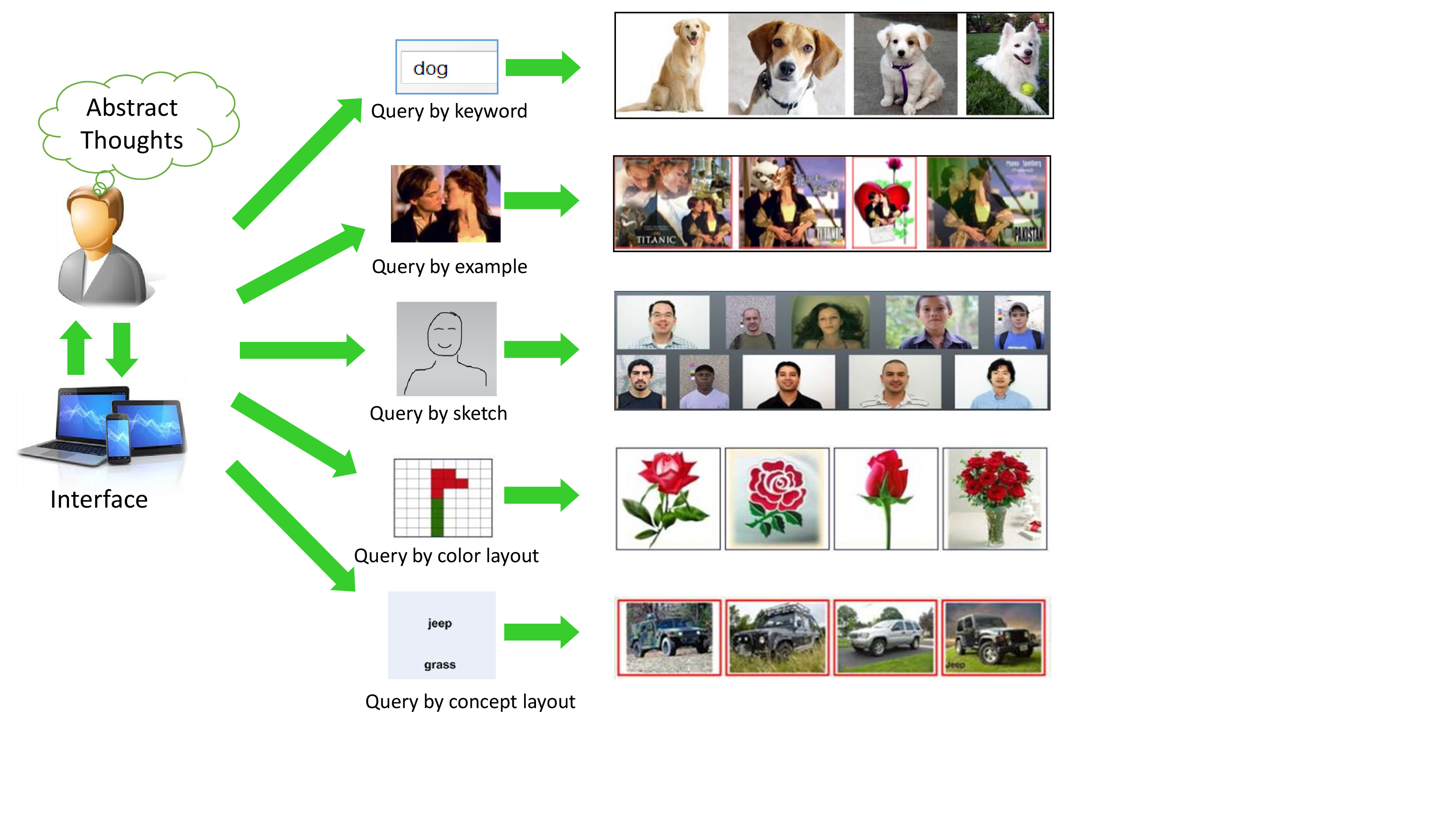}\\
  \caption{Illustration of different query schemes with the corresponding retrieval results.}\label{fig:query}
\end{figure*}

At the beginning of image retrieval, a user expresses his or her imaginary intention into some concrete visual query. The quality of the query has a significant impact on the retrieval results. A good and specific query may sufficiently reduce the retrieval difficulty and lead to satisfactory retrieval results. Generally, there are several kinds of query formation, such as query by example image, query by sketch map, query by color map, query by context map, \emph{etc}. As illustrated in Fig.~\ref{fig:query}, different query schemes lead to significantly distinguishing results. In the following, we will discuss each of those representative query formations.

The most intuitive query formation is query by example image. That is, a user has an example image at hand and would like to retrieve more or better images about the same or similar semantics. For instance, a picture holder may want to check whether his picture is used in some web pages without his permission; a cybercop may want to check a terrorist logo appearing in the Web images or videos for anti-terrorism. To eliminate the effect of the background, a bounding box may be specified in the example image to constrain the region of interest for query. Since the example images are objective without little human involvement, it is convenient to make quantitative analysis based on it so as to guide the design of the corresponding algorithms. Therefore, query by example is the most widely explored query formation style in the research on content-based image retrieval~\cite{sivic2003video}~\cite{nister2006scalable}~\cite{jegou2010improving}~\cite{zhou2012scalar}.

Besides query by example, a user may also express his intention with a sketch map~\cite{cao2010mindfind}~\cite{xiao2015sketch}. In this way, the query is a contour image. Since sketch is more close to the semantic representation, it tends to help retrieve target results in users' mind from the semantic perspective~\cite{cao2010mindfind}. Initial works on sketch based retrieval are limited to search for special artworks, such as clip arts~\cite{sousa2010sketch}~\cite{fonseca2009sketch} and simple patterns~\cite{liang2008sketch}. As a milestone, the representative work on sketch-based  retrieval for natural images is the edgel~\cite{cao2011edgel}. Sketch has also been employed in some image search engines, such as Gazopa\footnote{http://www.gazopa.com/} and Retrievr\footnote{http://labs.systemone.at/retrievr}. However, there are two non-trivial issues on sketch based query. Firstly, although some simple concepts, such as sun, fish, and flower, can be easily interpreted as simple shapes, in most time, it is difficult for a user to quickly sketch out what he wants to search. Secondly, since the images in the database are usually natural images, it needs to design special algorithms to convert them to sketch maps consistent with user intention.

Another query formation is color map. A user is allowed to specify the spatial distribution of colors in a given grid-like palette to generate a color map, which is used as query to retrieve images with similar colors in the relative regions of the image plain~\cite{wang2011interactive}. With coarse shape embedded, the color map based query can easily involve user interaction to improve the retrieval results but is limited by potential concepts to be represented. Besides, color or illumination change is prevalent in image capturing, which casts severe challenge on the reliance of color-based feature.

The above query formations are convenient for uses to input but may still be difficult to express the user's semantic intention. To alleviate this problem, Xu \emph{et al.} proposed to form the query with concepts by text words in some specific layout in the image plain~\cite{xu2010image}~\cite{xu2010interactive}. Such structured object query is also explored in~\cite{lan2012image} with a latent ranking SVM model. This kind of query is specially suitable for searching generalized objects or scenes with context when the object recognition results are ready for the database images and the queries.

It is notable that, in the above query schemes taken by most existing work, the query takes the form of single image, which may be insufficient to reflect user intension in some situations. If provided with multiple probe images as query, some new strategies are expected to collaboratively represent the the query or fuse the retrieval results of each single probe~\cite{kim2015ranking}. That may be an interesting research topic especially in the case of video retrieval where the query a video shot of temporal sequence.

\section{Image Representation}
In content based image retrieval, the key problem is how to efficiently measure the similarity between images. Since the visual objects or scenes may undergo various changes or transformations, it is infeasible to directly compare images at pixel level. Usually, visual features are extracted from images and subsequently transformed into a fix-sized vector for image representation. Considering the contradiction between large scale image database and the requirement for efficient query response, it is necessary to ``pack'' the visual features to facilitate the following indexing and image comparison. To achieve this goal, quantization with visual codebook training are used as a routine encoding processing for feature aggregation/pooling. Besides, as an important characteristic for visual data, spatial context is demonstrated vital to improve the distinctiveness of visual representation.

Based on the above discussion, we can mathematically formulate the content similarity between two images $\mathcal{X}$ and $\mathcal{Y}$ in Eq.~\ref{eq:ImgSimilarity}.

\begin{eqnarray}\label{eq:ImgSimilarity}
  S(\mathcal{X}, \mathcal{Y}) & = & \sum_{x \in \mathcal{X}}\sum_{y \in \mathcal{Y}}{k(x, y)} \\
                              & = & \sum_{x \in \mathcal{X}}\sum_{y \in \mathcal{Y}}{\phi(x)^{T}\phi(y)} \\
                              & = & \Psi(\mathcal{X})^{T}\Psi(\mathcal{Y}).
\end{eqnarray}

Based on Eq.~\ref{eq:ImgSimilarity}, there emerge three questions.
\begin{enumerate}
  \item Firstly, how to describe the content image $\mathcal{X}$ by a set of visual features $\{x_1, x_2, \cdots\}$?
  \item Secondly, how to transform feature sets $\mathcal{X} = \{x_1, x_2, \cdots\}$ with various sizes to a fixed-length vector $\Psi(\mathcal{X})$?
  \item Thirdly, how to efficiently compute the similarity between the fixed-length vectors $\Psi(\mathcal{X})^{T}\Psi(\mathcal{Y})$?
\end{enumerate}

The above three questions essentially correspond to the feature extraction, feature encoding \& aggregation, and database indexing, respectively. As for feature encoding and aggregation, it involves visual codebook learning, spatial context embedding, and quantization. In this section, we discuss the related works on those key issues in image representation, including feature extraction, visual codebook learning, spatial context embedding, quantization, and feature aggregation. The database indexing is left to the next section for discussion.

\subsection{Feature Extraction}
Traditionally, visual features are heuristically designed and can be categorized into local features and global features. Besides those hand-crafted features, recent years have witnessed the development of learning-based features. In the following, we will discuss those two kinds of features, respectively.

\subsubsection{Hand Crafted Feature}
In early CBIR algorithms and systems, global features are commonly used to describe image content by color~\cite{wengert2011bag}~\cite{wang2011interactive}, shape~\cite{cao2011edgel}~\cite{xie2015deepshape}~\cite{wang2015sketch}~\cite{bai2016gift}, texture~\cite{park2002fast}\cite{wang2014content}, and structure~\cite{wang2006large} into a single holistic representation. As one of the representative global feature, GIST feature~\cite{siagian2007rapid} is biologically plausible with low computational complexity and has been widely applied to evaluate approximate nearest neighbor search algorithms~\cite{kulis2009kernelized}~\cite{weiss2009spectral}~\cite{jegou2011product}~\cite{torralba2008small}. With compact representation and efficient implementation, global visual feature are very suitable for duplicate detection in large-scale image database~\cite{wang2006large}, but may not work well when the target images involve background clutter. Typically, global features can be used as a complementary part to improve the accuracy on near-duplicate image search based on local features~\cite{zhang2012query}.

Since the introduction of SIFT feature by Lowe~\cite{lowe1999object}~\cite{Lowe:IJCV04}, local feature has been extensively explored as a routine image representation in many works on content-based image retrieval. Generally, local feature extraction involves two key steps, \emph{i.e.} interest point detection and local region description. In interest point detection, some key points or regions with characteristic scale are detected with high repeatability. The repeatability here means that the interest points can be identified under various transformations or changes. Popular detectors include Difference of Gaussian (DoG)~\cite{Lowe:IJCV04}, MSER~\cite{matas2004robust}, Hessian affine detector~\cite{mikolajczyk2004scale}, Harris-Hessian detector~\cite{xie2011efficient}, and FAST ~\cite{Rosten2010faster}. In interest point detection, the invariance to translation and resizing is achieved. Distinguished from the above methods, it is also possible to obtain the interest points by uniformly and densely sample the image plane without any explicit detector~\cite{krizhevsky2011using}.

After the detection of interest points, a descriptor or multiple descriptors~\cite{wu2009multi} are extracted to describe the visual appearance of the local region centered at the interest point. Usually, the descriptor is designed to be invariant to rotation change and robust to affine distortion, addition of noise, and illumination changes, \emph{etc}. Besides, it should also be distinctive so as to correctly match a single feature with high probability against a large corpus of features from many images. Such property is especially emphasized in the scenario of large-scale visual applications. The most popular choice with the above merits is SIFT feature~\cite{Lowe:IJCV04}. As a variant, SURF~\cite{bay2006surf} is demonstrated with comparable performance but better efficiency.

Some improvements or extensions have been made on the basis of SIFT. In \cite{arandjelovic2012three}, Arandjelovic \emph{et al} proposed a root-SIFT by making root-normalization on the original SIFT descriptor. Although such operation is simple, it is demonstrated to significantly improve the image retrieval accuracy and can be readily plugged into many SIFT based image retrieval algorithms~\cite{zheng2014packing}. Zhou \emph{et al.} proposed to generate binary signature of the SIFT descriptor with two median thresholds determined by the original descriptor itself~\cite{zhou2012scalar}. The obtained binary SIFT leads to a new indexing scheme for image retrieval~\cite{zhou2015bsift}. Liu~\emph{et al.} extend the binary SIFT by first generating a binary comparison matrix via dimension-pair comparison and then flexibly dividing the matrix entries into segments each of which is hashed to a bit~\cite{liu2014cross}. In~\cite{zhang2012qsrank}, the SIFT descriptor is transformed to binary code with principal component analysis (PCA) and simple thresholding operations simply based on coefficients' sign. In~\cite{yu2011asift}, Affine-SIFT (ASIFT) simulates a set of sample views of the initial images by varying the two camera axis orientation parameters, \emph{i.e.}, the latitude and the longitude angles and covers effectively all six parameters of the affine transformation, consequently achieving fully affine invariance.

SIFT features extracted in regions with weak internal structure suffers poor distinctiveness and may degrade image retrieval performance. To identify and remove those features, Dong~\emph{et al.} regarded a SIFT descriptor as 128 samples of a discrete random variable ranging from 0 to 255 and make use of the entropy as a measurement metric to filter SIFT features with low entropy~\cite{dong2012high}.

Apart from floating point feature like SIFT, binary features are popularly explored and directly extracted from the local region of interest. Recently, binary feature BRIEF~\cite{BRIEF} and its variants, such as ORB~\cite{ORB}, FREAK~\cite{FREAK}, and BRISK~\cite{BRISK}, have been proposed and have attracted a great deal of attention in visual matching applications. Those binary features are computed by some simple intensity difference tests, which are extremely computationally efficient. With the advantage in efficiency from Hamming distance computation, those binary features based on FAST detector~\cite{Rosten2010faster} may have potential in large scale image search. In~\cite{zhang2015usb}, Zhang \emph{et al.} proposed a novel ultra short binary descriptor (USB) from the local regions of regions detected by DoG detector. The USB achieves fast image matching and indexing. Besides, following the binary SIFT scheme~\cite{zhou2012scalar}, it avoids the expensive codebook training and feature quantization in BoW model for image retrieval. A comprehensive evaluation of binary descriptors are referred in~\cite{madeo2016fast}.

Besides the gradient information in the local regions as in SIFT feature, edge and color can also be expressed into a compact descriptor, generating Edge-SIFT~\cite{zhang2013edge} and color-SIFT~\cite{van2010evaluating}. As a binary local feature, Edge-SIFT~\cite{zhang2013edge} describes a local region with the extracted Canny edge detection results. Zheng~\emph{et al} extracted color name feature from the local regions, which is further transformed to a binary signature to enhance the discrimination of local SIFT feature~\cite{zheng2014packing}.

\subsubsection{Learning-based Feature}

Apart from the above handcrafted visual features, it is also possible to learn features in a data-driven manner for image retrieval. Attribute feature, originally used for object categorization, can be used to represent the semantic characteristics for image retrieval~\cite{douze2011combining}~\cite{zhao2014affective}~\cite{tao2015attributes}. Generally, the attribute vocabulary can be manually defined by humans~\cite{farhadi2009describing}~\cite{khan2012color} or some ontology~\cite{torresani2010efficient}. For each attribute, a classifier can be trained with kernel over multiple low-level visual features based on labeled training image set and used to predict the attribute score for unseen images~\cite{torresani2010efficient}~\cite{khan2012color}~\cite{deng2011hierarchical}~\cite{cai2015attribute}. In ~\cite{zhang2013semantic}, the attribute feature is adopted as a semantic-aware representation to compensate local SIFT feature for image search. Karayev \emph{et al.} learned classifiers to predict image styles and applied it to search and rank image collection by styles~\cite{karayev2014recognizing}. The advantage of attribute feature is that it provides an elegant way to approximate the visual semantics so as to reduce the semantic gap. However, there are two issues on attribute features. Firstly, it is difficult to define a complete set of attribute vocabulary, either manually or in an automatic manner. Thus, the representation with the limited attribute vocabulary may be biased for a large and semantically diverse image database. Secondly, it is usually computationally expensive to extract attribute features due to the necessity to do classification over thousands of attribute categories~\cite{douze2011combining}~\cite{torresani2010efficient}.

Topic models, such as probabilistic Latent Semantic Analysis (pLSA) model~\cite{hofmann2001unsupervised} and Latent Dirichlet Allocation (LDA) model~\cite{blei2003latent}, are popularly adopted to learn feature representation with semantics embedded for image retrieval~\cite{horster2007image}~\cite{lienhart2007plsa}.

With the explosive research on deep neural network (DNN)~\cite{krizhevsky2011using}~\cite{simonyan2014very}~\cite{szegedy2014going}, recent years have witnessed the success of the learning-based features in multiple areas. With the deep architectures, high-level abstractions close to human cognition process can be learned~\cite{bengio2009learning}. As a result, it is feasible to adopt DNN to extract semantic-aware features by the activations of different lays in the networks. In~\cite{horster2008deep}, features are extracted in local patches with a deep restricted Boltzmann machine (DBN) which is refined by using back-propagation. As a typical structure of the DNN family, deep convolutional neural network (CNN)~\cite{krizhevsky2012imagenet} has demonstrated state-of-the-art performance in various tasks on image recognition and retrieval~\cite{Sharif2014CNN}. In~\cite{wan2014deep}, comprehensive studies is conducted on the potential of learned visual features with deep CNN for various applications including content based image retrieval. Razavian \emph{et al.} study the Alex-Net~\cite{krizhevsky2012imagenet} and VGG-Net~\cite{simonyan2014very}, and exploit the last convolutional layers response with max pooling as image representation for image retrieval~\cite{razavian2014visual}. In~\cite{zheng2015query}, the activations of the sixth layer of the Alex-Net~\cite{krizhevsky2012imagenet} is taken out as a DNN feature for each image, which is fused in the image similarity score level with traditional visual features including SIFT-based BoW feature, HSV histogram, and GIST.

Besides working as a global description of images, learning-based feature can also be obtained in a manner similar to local features~\cite{xie2015one}. The local regions of interest are generated by unsupervised object detection algorithms, such as selective search~\cite{uijlings2013selective}, objectness~\cite{alexe2012measuring}, and binarized normed gradients (BING)~\cite{cheng2014bing}. Those algorithms generate a number of object proposals in the form of bounding boxes. Then, in each object proposal region, the learning-based feature can be extracted. In ~\cite{sun2015scalable}, Sun \emph{et al.} adopted the CNN model to extract features from local image regions detected by a general object detector~\cite{cheng2014bing}, and applied it for image retrieval and demonstrated impressive performance. Considering the fact that object detection is sensitive to rotation transformation, Xie~\emph{et al.} proposed to rotate the test image by four different angles and then conduct object detection. Object proposals with top detection scores are then selected to extract the deep CNN feature~\cite{krizhevsky2012imagenet}. Tolias \emph{et al.}
generate feature vector of regional maximum activation of convolutions (R-MAC) towards geometry-aware re-ranking~\cite{tolias2016particular}. To speedup the max-pooling operation, a novel approximation is proposed by extending the idea of integral images. In~\cite{gordo2016deep}, the R-MAC descriptor is extended by selecting regions with a region-of-interest (ROI) selector based on region proposal network~\cite{ren2015faster}.

In the above approaches, the learning-based feature is extracted with the deep learning model trained for classification task. As a result, the learned feature may not well reflect the visual content characteristics of retrieval images, which may result in limited retrieval performance. Therefore, it is preferred to train the deep learning model directly for the retrieval task, which, however, is difficult since the potential image category in retrieval is difficult to define or enumerated. To partially address this difficulty, Babenko~\emph{et al.} focus on landmark retrieval and fine-tune the pre-trained CNN model on ImageNet with the class corresponding to landmarks~\cite{babenko2014neural}. after the fine-tuning, promising performance improvement is witnessed on the retrieval datasets with similar visual statistics, such as the Oxford Building dataset~\cite{philbin2007object}. To get rid of the dependence on examples or class labels, Paulin~\emph{et al.} proposed to generate patch-level feature representation based on convolutional kernel networks in an unsupervised way~\cite{paulin2015local}. In~\cite{xia2014supervised}, the supervision takes the form of binary codes, which are obtained by decomposing the similarity matrix of training images. The resultant deep CNN model is therefore ready to generate binary codes for images in an end-to-end way. Further, Lai~\emph{et al.} propose deep neuron networks to hash images into short binary codes with optimization based on triplet ranking loss~\cite{lai2015simultaneous}. The resulted short binary codes for image representation enable efficient retrieval by Hamming distance and considerable gain in storage.

\subsection{Visual Codebook Learning}
Usually, hundreds or thousands of local features can be extracted from a single image. To achieve a compact representation, high dimensional local features are quantized to visual words of a pre-trained visual codebook, and based on the quantization results an image with a set of local features can be transformed to a fixed-length vector, by the Bag-of-Visual-Words model~\cite{sivic2003video}, VLAD~\cite{jegou2010aggregating}, or Fisher Vector~\cite{perronnin2010large}. To generate a visual codebook beforehand, the most intuitive way is by clustering the training feature samples with brute-force~\emph{k}-means~\cite{sivic2003video}~\cite{jegou2008hamming} and then regarding the clustering centers as visual words. Since the local feature dimension is high and the training sample corpus is large, it suffers extremely high computational complexity to train a large, say, million-scale or larger, visual codebook. To address this problem, an alternative to to adopt the hierarchical~\emph{k}-means~\cite{nister2006scalable}, which reduces the computational complexity from linear to logarithm for large size visual codebook generation.

In the standard~\emph{k}-means, the most computing overhead is consumed on the assignment of feature samples to the close cluster center vector, which is implemented by linearly comparing all cluster center vectors. That process can be speeded up by replacing the linear scan with approximate nearest neighbor search. With such observation, Philbin~\emph{et al.} proposed an approximate~\emph{k}-means algorithm by exploiting randomized $k$-D trees for fast assignment~\cite{philbin2007object}. Instead of using~\emph{k}-means to generate visual words, Li~\emph{et al.} generated hyper-spheres by randomly sampling seed feature points with a predefined radius~\cite{li2009efficient}. Then, those hyper-spheres with the seed features corresponds to the visual codebook. In~\cite{chu2014graph}, Chu \emph{et al.} proposed to build the visual vocabulary based on graph density. It measures the intra-word similarity by the feature graph density and derives the visual word by dense feature graph with a Scalable Maximization Estimation (SME) scheme.

In the Bag-of-Visual-Words model, the visual codebook works as a media to identify the visual word ID, which can be regarded as the quantization or hashing result. In other words, it is feasible to directly transform the visual feature to a visual word ID without explicitly defining the visual word. Following this idea, different from the above codebook generation methods, some other approaches on image retrieval generate a virtual visual codebook without explicit training. Those methods transform a local feature to binary signature, based on which the visual word ID is heuristically defined. In~\cite{zhang2012qsrank}, Zhang \emph{et al.} proposed a new query-sensitive ranking algorithm to rank PCA-based binary hash codes to search for $\epsilon$-neighbors for image retrieval. The binary signature is generated with a LSH (locality sensitive hashing) strategy and the top bits are used as the visual word ID to group feature points with the same ID. Zhou \emph{et al.}~\cite{zhou2012scalar} proposed to binarize a SIFT descriptor into a 256-bit binary signature. Without training a codebook, this method selects 32 bits from the 256-bit vector as a codeword for indexing and search. The drawback of this approach is that the rest 224 bits per feature have to be stored in the inverted index lists, which casts a heavy overhead on memory. Similarly, Dong \emph{et al} proposed to transform to a SIFT descriptor to a 128-bit vector~\cite{dong2012high} with a sketch embedding technique~\cite{dong2008efficiently}. Then, the 128-bit vector is divided into 4 non-overlapped block, each of which is considered as a key or a visual word for later indexing. In~\cite{zhou2014towards}, Zhou \emph{et al} proposed a codebook-training-free framework based on scalable cascaded hashing. To ensure the recall rate of feature matching, the scalable cascaded hashing (SCH) scheme which conducts scalar quantization on the principal components of local descriptors in a cascaded manner.

\subsection{Spatial Context Embedding}
As the representation of structured visual content, visual features are correlated by spatial context in terms of orientation, scale, and key points' distance in image plane. By including the contextual information, the discriminative capability of visual codebook can be greatly enhanced~\cite{zhou2010BiSpace}. Analogy to the text phrase in information retrieval, it is feasible to generate visual phrase over visual words.  In~\cite{zhang2009descriptive}~\cite{zhang2011generating}, neighboring local features are correlated to generate high-order visual phrases, which are further refined to be more descriptive for content representation.

Many algorithms target on modeling the local spatial context among local visual features. Loose spatial consistency from some spatially nearest neighbors can be imposed to filter false visual-word matches. Supports are collected by checking the matched features with the search area defined by 15 nearest neighbors~\cite{sivic2003video}. Such loose scheme, although efficient, is sensitive to the image noise incurred by editing. Zhang~\emph{et al.} generated contextual visual codebook by modeling the spatial context of local features in group with a discriminant group distance metric~\cite{zhang2010building}. Wang~\emph{et al.} proposed descriptor contextual weighting (DCW) and spatial contextual weighting (SCW) of local features in the descriptor domain and spatial domain, respectively, to upgrade the vocabulary tree based approach~\cite{wang2011contextual}. DCW down-weights less informative features based on frequencies of descriptor quantization paths on a vocabulary tree while SCW exploits some efficient spatial contextual statistics to preserve the rich descriptive information of local features. In~\cite{Liu2012Embed}, Liu~\emph{et al.} built a spatial-relationship dictionary by embedding spatial context among local features for image retrieval.

Further, the multi-modal property that multiple different features are extracted at an identical key points is discussed and explored for contextual hashing~\cite{liu2014contextual}. In~\cite{chum2009geometric}, geometric min-hashing constructs repeatable hash keys with loosely local geometric information for more discriminative description. In~\cite{wu2009bundling}, Wu~\emph{et al.} proposed to bundle local features in a MSER region~\cite{matas2004robust}. The MSER regions are defined by an extremal property of the intensity function in the region and on its outer boundary and are detected as stable regions across a threshold range from watershed-based segmentation~\cite{matas2004robust}. Bundled features are compared by the shared visual word amount and the relative ordering of matched visual words. In~\cite{xie2011efficient}, ordinal measure (OM) feature~\cite{bhat1998ordinal} is extracted from the spatial neighborhood around local interest points. Then, local spatial consistency verification is conducted by checking whether the OMs of the correspondence features are below a predefined threshold.

Different from the above approaches, Cao~\emph{et al.} modeled the global spatial context by two families of ordered bag-of-features as a generation of the spatial pyramid matching~\cite{lazebnik2006beyond} by linear projection and circular projection and further refined them to capture the invariance of object translation, rotation, and scaling by simple histogram operations, including calibration, equalization, and decomposition~\cite{cao2010spatial}.

In the scenario of face retrieval, the above general codebook generation methods are likely to fail to capture the unique facial characteristics. To generate discriminative visual codebook, Wu~\emph{et al.} proposed to generate identity-based visual vocabulary with some training persons each with multiple face examples under various poses, expressions, and illumination conditions~\cite{wu2011scalable}. A visual word is defined as a tuple consisting of two components, \emph{i.e.}, person ID and position ID and associated with multiple examples.

\subsection{Feature Quantization}
With visual codebook defined, feature quantization is to assign a visual word ID to each feature. To design a suitable assignment function, special consideration should be made to balance quantization accuracy, efficiency, and memory overhead.

The most naive choice is to take the nearest neighbor search, so as to find the closest (the most similar) visual word of a given feature by linear scan, which, however, suffers expensive computational cost. Usually, approximate nearest neighbor (ANN) search methods are adopted to speed up the searching process, with sacrifice of accuracy to some extent. In~\cite{Lowe:IJCV04}, a~\emph{k}-d tree structure~\cite{bentley1990k} is utilized with a best-bin-first modification to find approximate nearest neighbors to the descriptor vector of the query. In~\cite{nister2006scalable}, based on the hierarchical vocabulary tree, an efficient approximate nearest neighbor search is achieved by propagating the query feature vector from the root node down the tree by comparing the corresponding child nodes and choosing the closest one.  In~\cite{Chanop}, a~\emph{k}-d forest approximation algorithm is proposed with reduced time complexity.  Muja and Lowe proposed a novel priority search k-means tree algorithm for scalable nearest neighbor search~\cite{Muja_flann_pami_2014} with FLANN library\footnote{http://www.cs.ubc.ca/research/flann/} provided. In~\cite{li2009efficient}, the feature quantization is achieved by range-based neighbor search over the random seeding codebook. This random seeding approach, although efficient in implementation, suffers the bias to the training data and achieves limited retrieval accuracy in large-scale image retrieval~\cite{zhou2016scalable}. Those approaches conduct quantization in a hard manner and inevitably incur severe quantization loss.

Considering that the codebook partitions the feature space into some non-overlapping cells, feature quantization works to identify which cell a test feature falls into. When the codebook size is large which means the feature space is finely partitioned, features proximate to the partition boundary are likely to fall into different cells. On the other hand, with small codebook and feature space coarsely partitioned, irrelevant features with large distance may also fall into the same cell. Both cases will incur quantization loss and degrade the recall and precision of feature matching, respectively. A trade-off shall be made on the codebook size to balance the recall and precision from the above two kinds of loss~\cite{nister2006scalable}, or some constraints are involved to refine the quantization quality.

Some approaches adopt a large visual codebook but take account of the soft quantization to reduce the quantization loss. Generally, a descriptor-dependent soft assignment scheme~\cite{philbin2008lost} is used to map a feature vector to a weighted combination of multiple visual words. Intuitively, the soft quantization can be performed for both a query feature and the database features. However, it will cost several times more memory to store the multiple quantization results for each database feature. As a trade-off, the soft quantization can be constrained to only the query side~\cite{jegou2010improving}. In~\cite{jegou2010improving}, a new quantizer is designed based on a codebook learned by brute-force $k$-means clustering. It first performs $k$-means clustering on the pre-trained visual words and generate a two-layer visual vocabulary tree in a bottom-up way. Then, new connections between the two-layer nodes are constructed by quantizing a large feature set with both layers of quantizers. Soft assignment is performed with a criteria based on distance ratio.

On the other hand, some other approaches keep a relatively small visual codebook but performs further verification to reduce the quantization loss. In~\cite{jegou2008hamming}, Hamming Embedding reduces the dimension of SIFT descriptors quantized to a visual word, and then trains a median vector by taking the median value in each dimension of the feature samples. After a new feature is quantized to a visual word, it is projected to the low dimensional space, and then compared with the median vector dimension-wise to generate binary signature for matching verification~\cite{wang2006large}. In~\cite{jain2011asymmetric}, a variant, \emph{i.e.}, the asymmetric Hamming Embedding scheme, is proposed to exploit the rich information conveyed by the binary signature. Zhou \emph{et al.}adopt a similar verification idea with a different binary signature which is generated by comparing each element of a feature descriptor to its median~\cite{zhou2013visual}.

The above approaches rely on single visual codebook for feature quantization. To correct quantization artifacts and improve recall, typically, multiple vocabularies are generated for feature quantization to improve
the recall~\cite{xia2013joint}\cite{jegou2012negative}. Since multiple vocabularies suffers from vocabulary correlation, Zheng \emph{et al} proposed a Bayes merging approach to down-weight the indexed features in the intersection set of multiple vocabularies~\cite{zheng2014bayes}. It models the the correlation problem in a probabilistic view and estimate a joint similarity on both image- and feature-level for the indexed features in the intersection set.

The vector quantization of local descriptors is closely related to approximate nearest neighbor search~\cite{jegou2011product}. In literature, there are many hashing algorithms for approximate nearest neighbor (ANN) search, such as LSH~\cite{indyk1998approximate}\cite{andoni2006near}, multi-probe LSH~\cite{lv2007multi}, kernelized LSH~\cite{kulis2009kernelized}, semi-supervised hashing method (SSH)~\cite{wang2010semi}, spectral hashing~\cite{weiss2009spectral}, min-Hashing~\cite{chum2008near}, iterative quantization~\cite{gong2011iterative}, random grids~\cite{aigerrandom}, bucket distance hashing (BDH)~\cite{iwamuramost}, query-driven iterated neighborhood graph search~\cite{wang2012query}, and linear distance preserving hashing~\cite{wang2016linear}. These hashing methods, however, are mostly applied to global image features such as GIST or BoW features at the image level, or to feature retrieval only at the local feature level. There is few work dedicated to image level search based on local feature hashing~\cite{he2012mobile}. The major concern of those hashing methods is that multiple hashing tables are usually involved and each feature needs to be indexed multiple times, which cast heavy memory burden. Besides, in hashing methods such as LSH~\cite{andoni2006near}, multi-probe LSH~\cite{lv2007multi} and kernelized LSH~\cite{kulis2009kernelized}, the original database feature vectors need be kept in memory to compute the exact distance to the query feature, which is infeasible in the scenario of large-scale image search with local features. Moreover, approximate nearest neighbor search usually targets at identifying the top-\emph{k} closest data to the query, which ignores the essence of range-based neighbor search in visual feature matching. That is, given a query feature, the number of target data in the database is query-sensitive and determined by the coverage of the range-based neighborhood of the query.

In~\cite{jegou2011product}, a novel product quantization is proposed to generate an exponentially large codebook with low cost in memory and time for approximate nearest neighbor search. It decomposes the feature space into a Cartesian product of low-dimensional subspaces and quantizes each sub-space individually. The quantization indices of each sub-space are presented as a short code, based on which the Euclidean distance between two feature vectors can be efficiently estimated by looking up a pre-computed table. The product quantization, however, suffers from exhaustive search for identifying target features, which is prohibitive in large-scale image search~\cite{jegou2011product}. As a partial solution to this bottle neck, vector quantization by \emph{k}-means can be involved to narrow the search scope and allow the product to focus on a small fraction of indexed features~\cite{jegou2011product}. In~\cite{ge2013optimized}, the product quantization is optimized with respect to the vector space decomposition and the quantization codebook with two solutions from the non-parametric and the parametric perspectives. Zhou \emph{et al.} formulated the feature matching as an $\epsilon-$neighborhood problem and approximated it with a dual-resolution quantization scheme for efficient indexing and querying~\cite{zhou2016scalable}. It performs scalar quantization in coarse and fine resolutions on each dimension of the data, and cascades the quantization results over all dimensions. The cascaded quantization results in coarse resolution are used to build the index, while the cascaded quantization results in fine resolutions are transformed to a binary signature for matching verification.

In~\cite{tuytelaars2007vector}, the high dimensional SIFT descriptor space is partitioned into regular lattices. Although demonstrated to work well in image classification, in~\cite{philbin2008lost}, regular lattice quantization is revealed to work much worse than~\cite{nister2006scalable}~\cite{philbin2008lost} in large scale image search application.

\subsection{Feature Aggregation}
When an image is represented by a set of local features, it is necessary to aggregate those local features into a fixed-length vector representation for convenience of similarity comparison between query and database images. Generally, there are three alternatives to achieve this goal. The first one is the classic Bag-of-Visual-Words representation, which quantizes each local feature to the closest visual word of a pre-trained visual codebook. The quantization result of a single local feature can be regarded as a high-dimensional binary vector, where the non-zero dimension corresponds to the quantized visual word. By pooling the quantization results of all local features in an image, we obtain a BoW vector with the dimension size as the visual codebook size. In this scheme, the involved visual codebook is usually very large in size and the generated BoW vector is very sparse, which facilitates the use of the inverted file indexing.

The second popular feature aggregation method is the VLAD (vector of locally aggregated descriptors) approach~\cite{jegou2010aggregating}, which adopts k-means based vector quantization and accumulates the quantization residues for features quantized to each visual word and concatenate those accumulated vectors into a single vector representation. With compact size, the VLAD vector inherits some important properties from SIFT feature, including invariance to translation, rotation, and scaling.  In~\cite{arandjelovic2013all}, the VLAD approach is improved by a new intra-normalization scheme and multiple spatial VLAD representation. An in-depth analysis on VLAD is conducted in~\cite{spyromitros2014comprehensive}. In~\cite{jegou2014triangulation}, an extension of VLAD is proposed with triangulation embedding scheme and democratic aggregation technique. Further, Tolias~\emph{et al.} encompassed the VLAD vector with various matching schemes~\cite{Tolias2013to}. To reduce the computational complexity of the democratic aggregation scheme, Gao \emph{et al.} proposed a fast scheme with comparable retrieval accuracy performance~\cite{gao2015fast}. In~\cite{ge2013sparse}, sparse coding is adopted to encode the local feature descriptors into sparse vectors, which are further aggregated with a max-pooling strategy. Liu \emph{et al.} proposed a hierarchical scheme to build the VLAD vector with SIFT feature~\cite{liu2015uniforming}. By involving a hidden-layer vocabulary, the distribution of the residue vectors to be aggregated becomes much more uniform, leading to better discrimination for the representation.

Although compact and efficient representation is achieved by global aggregation of all local features in an image, the original VLAD vector sacrifices the flexibility to address partial occlusion and background clutter. To alleviate this problem, Liu \emph{et al.}~\cite{liu2014uniting} grouped local key points by their spatial positions in the image plane and aggregated all local descriptors in each group by the VLAD scheme~\cite{jegou2010aggregating}. As a result, a local aggregation of local features is achieved and promising retrieval accuracy is demonstrated with a tradeoff in memory cost.

Besides the BoW representation and the VLAD, another alternative is the Fisher Vector based representation~\cite{perronnin2010large} with Fisher kernel~\cite{Jaakkola1998fisher}~\cite{jaakkola1999exploiting}. As a generative model, given a set of features for an image, Fisher vector represents them into a fix-sized vector by the gradient of the log-likelihood function with respect to a set of parameter vectors~\cite{sanchez2013image}. In~\cite{perronnin2010large}~\cite{duan2013compact}, Gaussian Mixture Model (GMM) is adopted as a generative model to aggregate the normalized concatenated gradient vectors of all local descriptors into a uniform Fisher vector with an average pooling scheme. In fact, the Fisher Vector can be regarded as a generalized representation of the BoW representation and VLAD. On one hand, if we keep only the gradient of the log-likelihood function with respect to the weight of GMM, the Fisher Vector degenerates to a soft version of the BoW vector. On the other hand, If we keep only the gradient of the log-likelyhood function with respect to the mean vector of GMM, we can derive the VLAD representation~\cite{jegou2011product}.

In either the Fish vector or VLAD representation, the involved GMM number or codebook size is relative small and the obtained aggregated vector is no long sparse. As a result, it is unsuitable to apply the inverted file indexing scheme to index images based on the aggregated results. To address this dilemma, the aggregated vector is dimensionally reduced and further encoded by product quantization~\cite{jegou2011product} for efficient distance computation.

The above aggregation schemes are based on local hand-crafted feature, such as SIFT feature. Intuitively, such schemes can be directly leveraged to local deep features. Following this idea, Gong \emph{et al.}~\cite{gong2014multi} extract local CNN features from the local patches sampled regularly at multiple scale levels and pool the CNN features in each scale level with the VLAD scheme [37]. In~\cite{babenko2015aggregating}, Babenko~\emph{et al.} interpret the activations from the last convolutional layers of CNNs as local deep features. They reveal that the individual similarity of local deep feature is very discriminative and the simple aggregation with sum pooling over local deep feature yields the best performance.

\section{Database Indexing}
Image index refers to a database organizing structure to assist for efficient retrieval of the target images. Since the response time is a key issue in retrieval, the significance of database indexing is becoming increasingly evident as the scale of image database on the Web explosively grows. Generally, in CBIR, two kinds of indexing techniques are popularly adopted, \emph{i.e.}, inverted file indexing and hashing based indexing. In the following, we will discuss related retrieval algorithms in each category, respectively.

\subsection{Inverted File Indexing}
Inspired by the success of text search engines, inverted file indexing~\cite{baeza1999modern} has been successfully used for large scale image search ~\cite{sivic2003video}~\cite{philbin2007object}~\cite{zhou2010spatial}~\cite{chum2007total}~\cite{nister2006scalable}~\cite{jegou2008hamming}~\cite{wu2009bundling}~\cite{cai2014scalable}. In essence, the inverted file structure is a compact column-wise representation of a sparse matrix, where the row and the column denote image and visual word, respectively. In on-line retrieval, only those images sharing common visual words with the query image need to be checked. Therefore, the number of candidate images to be compared is greatly reduced, achieving an efficient response.

In the inverted file structure, each visual word is followed by an inverted file list of entries. Each entry stores the ID of the image where the visual word appears, and some other clues for verification or similarity measurement. For instance, Hamming Embedding~\cite{jegou2008hamming} generates a 64-bit Hamming code for each feature to verify the descriptor matching. The geometric clues, such as feature position, scale, and orientation, are also stored in the inverted file list for geometric consistency verification ~\cite{philbin2007object}~\cite{zhou2010spatial}~\cite{jegou2008hamming}~\cite{zhou2011large}. In~\cite{wu2009bundling}, Wu \emph{et al.} recorded the feature orders in horizontal and verification direction in each bundled feature located in a MSER region. In~\cite{wang2011contextual}, 3 spatial statistics, including descriptor density, mean relative log scale, and mean orientation difference, are calculated for each feature and stored in the inverted list after quantization. Zheng~\emph{et al.} modeled the correlation between multiple features with a multi-IDF scheme and coupled the binary signatures of those features into the inverted file to enhances the quality of visual matching~\cite{zheng2014coupled}.

Following the general idea of inverted file structure, many variants are proposed. In~\cite{cao2011edgel}, to adapt to the inverted index structure for sketch-based retrieval, it regularly quantizes the edge pixel in position channel and orientation channel and follows each entry in the edgel dictionary with an inverted lists of related images. In~\cite{zheng2014packing}, Zheng \emph{et al} proposed a new coupled Multi-Index (c-MI) framework to fuse complementary features at indexing level. Each dimension of c-MI corresponds to one kind of feature, and the retrieval process votes for images similar in both SIFT and color attribute~\cite{khan2012color} feature spaces. In~\cite{liu2014cross}, the image database is cross-indexed in both the binary SIFT space and the original SIFT space. With such cross-indexing structure, a new searching strategy is designed to find target data for effective feature quantization.

Some methods try to embed the semantics into the index structure. In~\cite{zhang2009efficient}, Zhang \emph{et al} proposed a new indexing structure by decomposing a document-like representation of an image into two components, one for dimension reduction and the other for residual information preservation. The decomposition is achieved by either a graphical model or a matrix factorization approach.  Then, the similarity between images is transferred to measuring similarities of their components. In~\cite{zhang2013semantic}, Zhang \emph{et al} proposed a semantic-aware co-indexing to jointly embed two strong cues, \emph{i.e.}, local SIFT feature and semantic attributes, into the inverted indexes. It exploits 1000 semantic attributes to filter out isolated images and insert semantically similar images to the initial inverted index set built based on local SIFT features. As a result, the discriminative capability of the indexed features is significantly enhanced.

To adapt the product quantization~\cite{jegou2011product} to the inverted index idea, inverted multi-index is proposed to generalize the inverted index idea by replacing the standard quantization within inverted indices with product quantization, so as to speed up the approximate nearest neighbor search.

To improve the recall rate of inverted indexing algorithms, the database images are indexed multiple times with multiple quantizers, such as randomized k-d trees~\cite{silpa2008optimised}~\cite{wu2009multi}. In~\cite{xia2013joint}, a joint inverted indexing algorithm is proposed, which jointly optimizes all codewords in all quantizers and demonstrates considerable improvement over methods with multiple independent quantizers. In~\cite{arandjelovic2012three}, this goal is achieved by augmenting the image features for the database images which are estimated to be visible in a homograpy in the augmented images.

To speedup the online retrieval process, Zheng~\emph{et al.} proposed a novel Q-Index structure based on the inverted index organization~\cite{zheng2015fast}. It defines an impact score for each indexed local SIFT feature based on TF-IDF, scale, saliency, and quantization ambiguity. Then, based on the impact score, it introduced two complementary strategies, \emph{i.e.} query pruning and early termination, with the former to discard less important features in the query and the later to partially visit the index lists containing the most important indexed features. The proposed algorithm demonstrates significant speed-up for online query with competitive retrieval accuracy. In~\cite{ji2013learning}, Ji~\emph{et al.} considered the scenario of parallelized image retrieval and proposed to distribute visual indexing structure over multiple servers. To reduce the search latency across servers, it formulates the index distribution problem as a learning problem by maximizing the uniformity of assigning the words of a given query to multiple servers.

\subsection{Hashing Based Indexing}
When the image representation, for instance GIST feature and VLAD feature, is a dense vector with the majority of the coefficients being non-zero, it is unsuitable to directly apply the inverted file structure for indexing. To achieve efficient retrieval for relevant results, hashing techniques are popularly adopted~\cite{heo2012spherical}~\cite{tang2015neighborhood}~\cite{wu2015distance}~\cite{jiang2015revisiting}~\cite{liu2016deep}. The most representative hashing scheme is the locality sensitive hashing (LSH)~\cite{datar2004locality}, which partitions the feature space with multiple hash functions of random projections with the intuition that for objects which are close to each other, the collision probability is much higher than for those which are far away. Given a query, some candidates are first retrieved based on hashing collision and re-ranked based on the exact distance from the query. In~\cite{kulis2009kernelized}, LSH is generated to accommodate arbitrary kernel functions, with sub-linear time approximate similarity search permitted. The potential concern of those hashing scheme is that, since the raw database representation vectors should be stored in memory for the reranking stage, they are not well scalable to large-scale image database. In~\cite{avrithis2010feature}, a feature map is proposed by integrating appearance and global geometry, which is further hashed for indexing. This scheme, however, suffers expensive memory cost which is quadratic in the number of local features, which limits its scalability towards large scale image retrieval. To address this drawback, an extension is made with a feature selection model to replace the hashing approach~\cite{tolias2014towards}.

With the inverted index structure, the memory cost is proportional to the amount of non-zero elements in the representation vector. To further reduce such memory overhead, Jegou~\emph{et al.} proposed to approximate the original visual word occurrence vector by projecting it onto a set of pre-defined sparse projection functions, generating multiple min-BOF descriptors~\cite{jegou2009packing}. Those min-BOF descriptors is further quantized for indexing. With similar attempt, in~\cite{chum2008near}\cite{chum2007scalable}, min-Hash is proposed to describe images by mapping the visual word occurrence vector to a low-dimensional representation by a group of min-hash functions and define image similarity as the visual word set overlap. Consequently, only a small constant amount of data per image need to be stored. The potential concern of min-hashing~\cite{chum2008near}\cite{chum2007scalable} and its variant~\cite{chum2009geometric} is that although high retrieval precision can be achieved, the retrieval recall performance may be limited unless many more hashing tables are involved, which, however, imposes severe memory burden.


\section{Image Scoring}
In multimedia retrieval, the target results in the index image database are assigned with a relevance score for ranking and then returned to users. The relevance score can be defined either by measuring distance between the aggregated feature vectors of image representation or from the perspective of voting from relevant visual feature matches.

\subsection{Distance Based Scoring}
With feature aggregation, an image is represented into a fix-sized vector. The content relevance between images can be measured based on the $L_p$-normalized distance between their feature aggregation vectors, as shown in Eq.~\ref{eq:score}.

\begin{equation}
D(I_q, I_m) = \left(\sum^N_{i = 1} \left|q_i - m_i\right|^p\right)^{\frac{1}{p}}
\label{eq:score}
\end{equation}
where the feature aggregation vectors of image $I_q$ and $I_m$ are denoted as $[q_1, q_2, \cdots, q_N]$ and $[m_1, m_2, \cdots, m_N]$, respectively, and $N$ denotes the vector dimension. In~\cite{nister2006scalable}, it is revealed that $L_1$-norm yields better retrieval accuracy than $L_2$-norm with the BoW model. Lin~\emph{et al.} extended the above feature distance to measure partial similarity between images with an optimization scheme~\cite{lin2010local}.

When the BoW model is adopted for image representation, the feature aggregation vector is essentially a weighted visual word histogram obtained based on the feature quantization results. To distinguish the significance of visual words in different images, term frequency (TF) and inverted document/image frequency (IDF) are widely applied in many existing state-of-the-art algorithms~\cite{nister2006scalable}\cite{jegou2008hamming}\cite{sivic2003video}\cite{philbin2008lost}\cite{wu2009bundling}. Generally, the visual word vector weighted by TF and IDF are $L_p$-normalized for later distance computation. When the codebook size is much larger than the local feature amount in images, the aggregated feature vector of image is very sparse and we only need to check those visual words appearing in both images as illustrated in Eq.~\ref{eq:score2}~\cite{nister2006scalable}, which is very efficient in practical implementation.

\begin{eqnarray}
D(I_q, I_m) & = & \sum^N_{i = 1} |q_i - m_i|^p \\
      & = & 2 + \sum_{i | q_i \neq 0, m_i \neq 0}\left(|q_i - m_i|^p - q_i^p - m_i^p\right)
\label{eq:score2}
\end{eqnarray}

However, the dissimilarity measure by the $L_p$-distance is not optimal. As revealed in~\cite{jegou2010accurate}, there exists the neighborhood reversibility issue, which means that an image is usually not the $k$-nearest neighbor of its $k$-nearest neighbor images. Such issue causes that problem that some images are frequently returned while others are rarely returned when submitting query images. To address this problem, Jegou \emph{et al.} proposed a novel contextual dissimilarity measure to refine the Euclidean distance based distance~\cite{jegou2010accurate}. It modifies the neighborhood structure in the BoW space by iteratively estimating distance update terms in the spirit of Sinkhorns scaling algorithm. Alternatively, in~\cite{qin2013query}, a probabilistic framework is proposed to model the feature to feature similarity measure and a query adaptive similarity is derived. Different from the above approaches, in~\cite{donoser2013diffusion}, the similarity metric is implicitly learnt with diffusion processes by exploring the affinity graphs to capture the intrinsic manifold of database images.

In~\cite{jegou2012negative}, Jegou \emph{et al.} investigated the phenomenon of co-missing and co-occurrence in the regular BoW vector representation. The co-missing phenomenon denotes a negative evidence, \emph{i.e.}, a visual word is jointly missing from two BoW vectors. To include the under-estimated evidence for similarity measurement refinement, vectors of images are centered by mean substraction~\cite{jegou2012negative}. On the other hand, the co-occurrence of visual words across BoW vectors will cause over-counting of some visual patterns. To limit this impact, a whitening operation is introduced to the BoW vector to generate a new representation~\cite{jegou2012negative}. Such preprocessing also applies to the VLAD vector~\cite{jegou2010aggregating}. Considerable accuracy gain has been demonstrated with the above operations.

\subsection{Voting Based Scoring}
In local feature based image retrieval, the image similarity is intrinsically determined by the feature matches between images. Therefore, it is natural to derive the image similarity score by aggregating votes from the matched features. In this way, the similarity score is not necessarily normalized, which is acceptable considering the nature of visual ranking in image retrieval.

In~\cite{zhou2011large}, the relevance score is simply defined by counting how many pairs of local feature are matches across two images. In~\cite{jegou2010improving}, Jegou~\emph{et al} formulated the scoring function as a cumulation of squared TF-IDF weights on shared visual words, which is essentially a BOF (bag of features) inner product~\cite{jegou2010improving}. In~\cite{wu2009bundling}, the image similarity is defined as the sum of the TF-IDF score~\cite{zhang2011GVP}, which is further enhanced with a weighting term by matching bundled feature sets. The weighting term consists of membership term and geometric term. The former term is defined as the number of shared visual words between two bundled features, while the latter is formulated using relative ordering to penalize geometric inconsistency of the matching between two bundled features. In~\cite{zheng2013lp}\cite{zheng2014lp}, Zheng \emph{et al} propose a novel Lp-norm IDF to extend the classic IDF weighting scheme.

The context clues in the descriptor space and the spatial domain are important to contribute the similarity score when comparing images.  In~\cite{wang2011contextual}, a contextual weighting scheme is introduced to enhance the original IDF-based voting so as to improve the classic vocabulary tree approach. Two kinds of weighting scheme, \emph{i.e.}, descriptor contextual weighting (DCW) and spatial contextual weighting, are formulated to multiply the basic IDF weight as a new weighting scheme for image scoring. In~\cite{shen2012object}, Shen~\emph{et al.} proposed a spatially-constrained similarity measure based on a certain transformation to formulate voting score. The transformation space is discretized and a voting map is generated based on the relative locations of matched features to determine the optimal transformation.

In~\cite{jegou2009packing}, each indexed feature is embedded with a binary signature and the image distance is defined as a summation of the hamming distance between matched features, of which the distance for the unobserved match is set as statistical expectation of the distance. Similar scoring scheme for the unobserved match is also adopted by Liu \emph{et al.}~\cite{liu2014uniting}. In~\cite{xie2011efficient}, to tolerate the correspondences of multiple visual objects with different transformations, local similarity of deformations is derived from the peak value in the histogram of pairwise geometric consistency~\cite{xie2011pairwise}. This similarity score is used as a weighting term to the general voting scores from local correspondences.

In image retrieval with visual word representation, similar to text-based information retrieval~\cite{katz1996distribution}, there is a phenomenon of visual word burstiness, \emph{i.e.}, some visual element appears much more frequently in an image than the statistically expectation, which undermines the visual similarity measure. To address this problem, Jegou~\emph{et al} proposed three strategies to penalize the voting scores from the bursting visual words by removing multiple local matches and weaken the influence of intra- and inter-images bursts~\cite{jegou2009burstiness}~\cite{shi2015early}.

\section{Search Reranking}
The initially returned result list can be further refined by exploring the visual context~\cite{bai2016sparse, yang2015re} or enhancing the original query. Geometric verification~\cite{philbin2007object}~\cite{zhou2010spatial}~\cite{jegou2008hamming}~\cite{zhou2011large}~\cite{chum2009geometric}~\cite{li2015pairwise}, query expansion~\cite{chum2007total}~\cite{kuo2009query}, and retrieval fusion~\cite{zhang2012query} are three of the most successful post-processing techniques to boost the accuracy of large scale image search. In the following, we will review the related literature in each category.

\subsection{Geometric Context Verification}
In image retrieval with local invariant features, the feature correspondences between query and database images are built based on the proximity of local features in the descriptor space. As a popular criteria, a tentative correspondence is built if the corresponding two local features are quantized to the same visual word of a pre-trained visual vocabulary. However, due to the ambiguity of local descriptor and the quantization loss, false correspondences of irrelevant visual content are inevitably incurred, which confuse the similarity measurement for images and degrade the retrieval accuracy. Note that, besides the descriptor, local invariant features are characterised by other geometric context, such as the location of key points in image plane, orientation, scale, and spatial co-occurrences with other local features. Such geometric context is an important clue to depress or exclude those false matches.

Generally, among the inliers in the correspondences set, there is an underlying transformation model. If the model is uncovered, we can easily distinguish the inliers from the outliers. To model the transformation of visual object or scene across images, an affine transformation model with six parameters can be used, which estimates the rotation, scaling, translation, and perspective change in a single homography~\cite{philbin2007object}. For some difficult cases, there may exist multiple homographies which makes the model estimation problem much more challenging.

Some approaches estimate the transformation model in an explicit way to verify the local correspondences. Those methods are either based the RANSAC-like idea~\cite{philbin2007object}\cite{Lowe:IJCV04}\cite{chum2005matching}~\cite{xie2011efficient} or follow the Hough voting strategy~\cite{Lowe:IJCV04}\cite{Tolias2011hough}. The key idea of RANSAC~\cite{fischler1981random} is to generate hypotheses on random sets of correspondences and identify a geometric model with the maximum inliers. Statistically speaking, the genuine model can be recovered with sufficient number of correspondence sampling and model evaluation. However, when the rate of inliers is small, the expected number of correspondence sampling is large, which incurs high computational complexity. In~\cite{philbin2007object}, by adopting the region shape of local feature, a hypothesis is generated with single correspondence, which make it feasible to enumerate all hypotheses and significantly reduces the computational cost compared with RANSAC. There are two issues on the RANSAC based algorithms. Firstly, it needs a parameter for hypothesis verification, which is usually defined in an ad-hoc way. Secondly, the computational complexity is quadratic with respect to the number of correspondences, which is somewhat expensive.

An alternative to the RANSAC-like methods, Hough voting strategy~\cite{Lowe:IJCV04}~\cite{avrithis2014hough} operates in a transformation space. In this case, the voting operation is linear to the correspondence number. In~\cite{jegou2008hamming}, the Hough voting is conducted in the space of scale and orientation. Based on the SIFT feature correspondences between images, it builds two histograms on the orientation difference and scale difference separately. Assuming that truly matched features will share similar orientation difference, it identifies the peak points in the histogram on orientation difference of matched features and regard those feature pairs with orientation difference far from the peak as irrelevant and false matches. Similar operation is also performed on the scale difference of matched features to further remove those geometrically inconsistent SIFT matches. In~\cite{zhang2011GVP}, Zhang \emph{et al.} built a 2D Hough voting space based on the relative displacements of corresponding local features to derive the geometric-preserving visual phrase (GVP).This algorithm can be extended to address the transformation invariance to scale and rotation with the price of high memory overhead to maintain the Hough histograms. The potential problem in Hough voting is the flexibility issue in the definition of the bin size for the transformation space partition. To address the problem, in~\cite{Tolias2011hough}, motivated by the pyramid matching scheme~\cite{grauman2005pyramid}, Tolias \emph{et al.} propose a Hough pyramid matching scheme. It approximates affinity by bin size and group the correspondences based on the affinity in a bottom-up way. Notably, the complexity of this algorithm is linear to the correspondence number. In~\cite{avrithis2014hough}, the Hough pyramid matching scheme is extended by including the soft assignment for feature quantization on the query image. Different from the above methods, Li \emph{et al.} proposed a novel pairwise geometric matching method~\cite{li2015pairwise} for implicit spatial verification at a significantly reduced computational cost. To reduce the correspondence redundancy, it first builds the initial correspondence set with a one-versus-one matching strategy, which is further refined based on Hough voting in the scaling and rotation transformation space~\cite{jegou2008hamming}. Based on the reliable correspondence set, a new pairwise weighting method is proposed to measure the matching score between two images.

Some other algorithms approach the geometric context verification problem without explicit handling the transformation model. Sivic~\emph{et al.} adopted the consistency of spatial context in local feature groups to verify correspondences~\cite{sivic2003video}. In~\cite{zhou2010spatial}, a spatial coding scheme is proposed to encode into two binary maps by comparing the relative coordinates of matched feature points in horizontal and vertical directions, respectively. Then it recursively removes geometrically inconsistent matches by analyzing those maps. Although spatial coding map is invariant to image changes in translation and scaling, it cannot handle the rotation change. In~\cite{zhou2011large}~\cite{zhou2013sift}, Zhou \emph{et al.} extended the spatial coding by including the characteristic orientation and scale of SIFT feature and proposed two geometric context coding methods, \emph{i.e.}, geometric square coding and geometric fan coding. The geometric coding algorithm can well handle image changes in translation, rotation, and scaling. In~\cite{chu2013robust}, Chu~\emph{et al.} proposed a Combined-Orientation-Position (COP) consistency graph model to measure the relative spatial consistency among the candidate matches of SIFT features with a coarse-to-fine family of evenly sectored polar coordinate system. Those spatially inconsistent noisy features are effectively identified and rejected by detecting the group of candidate feature matches with the largest average COP consistency.

\subsection{Query Expansion}
Query expansion, leveraged from text retrieval, reissues the initially highly-ranked results to generate new queries. Some relevant features, which are not present in the original query, can be used to enrich the original query to further improve the recall performance. Several expansion strategies, such as average query expansion, transitive closure expansion, recursive average query expansion, intra-expansion, and inter-expansion, \emph{etc.}, have been discussed in~\cite{chum2007total}~\cite{kuo2009query}.

In~\cite{arandjelovic2012three}, a discriminative query expansion algorithm is proposed. It takes spatially verified images as positive data and images with low tf-idf scores as the negative training data. Then, a classifier is learnt on-the-fly and images are sorted by their signed distances from the decision boundary. In~\cite{xie2014fast}, Xie~\emph{et al.} constructed a sparse graph by connecting potentially relevant images offline and adopted a query-dependent algorithm, \emph{i.e.}, HITS~\cite{kleinberg1999authoritative}, to reranking images based on affinity propagation. Further, Xie~\emph{et al.} formulated the search process with a heterogeneous graph model and proposed two graph-based re-ranking algorithms to improve the
search precision and recall, respectively~\cite{xie2015heterogeneous}. It first incrementally identifies the most reliable images from the database to expand the query so as to boost the recall. After that, an image-feature voting scheme is used to iteratively update the scores of images and features to re-rank images. In~\cite{xie2014contextual}, a contextual query expansion scheme is proposed to explore the common visual patterns. The contextual query expansion is performed in both the visual word level and the image level.

As a special case of query expansion, relevance feedback~\cite{rui1998relevance} has been demonstrated to be successful search re-ranking technique and well studied before 2000 and received some attention in recent years~\cite{tao2004random}~\cite{tao2006asymmetric}~\cite{hoi2008semi}~\cite{arevalillo2013improved}~\cite{rabinovich2014utilizing}~\cite{wang2014image}. In relevance feedback, the key idea is to learn a query-specific similarity metric based on the relevant and irrelevant examples indicated by users. Some discriminative models are learned with SVM~\cite{tao2004random}\cite{tao2006asymmetric} or boosting schemes~\cite{tieu2004boosting}. Considering that users are usually reluctant or impatient to specify positive or negative images, user click log information can be collected as feedback to implicitly improve the retrieval system~\cite{zhang2013image}~\cite{yu2015learning}. For more discussion on relevance feedback, we refer readers to~\cite{zhou2003relevance}~\cite{patil2011relevance} for a comprehensive survey.

\subsection{Retrieval Fusion}
An image can be represented by different features, based on which different methods can be designed for retrieval. If the retrieval results of different methods are complementary to each other, they can be fused to obtain better results. Most approaches conduct retrieval fusion in the rank level. Fagin~\emph{et al.} proposed a rank aggregation algorithm to combine the image ranking lists of multiple independent retrieval methods or ``voters''~\cite{fagin2003efficient}. In~\cite{zhang2012query}, the retrieval fusion is formulated as a graph-based ranking problem. A weighted undirected graph is built based on the retrieval results of one method and the graphs corresponding to multiple retrieval methods are fused to a single graph, based on which, link analysis~\cite{page1999pagerank} or maximizing weighted density is conducted to identify the relevance score and rank the retrieval results. In~\cite{ye2012robust}, Ye~\emph{et al.} proposed a novel rank minimization method to fuse the confidence scores of multiple different models. It first constructs a comparative relationship matrix based on the predicted confident scores for each model. With the assumption that the relative score relations are consistent across different models with some sparse deviations, it formulates the score fusion problem as seeking a shred rank-2 matrix and derives a robust a score vector.

Different from the above fusion methods, Zheng \emph{et al.} approached the retrieval fusion in the score level~\cite{zheng2015query}. Motivated by the shape differences in the ranked score curve between good and bad representation features, it normalizes the score curves by reference curves trained on irrelevant data and derives an effectiveness score based on the area under the normalized score curve. Then, the query similarity measurement is adaptively formulated in a product manner over the feature scores weighted by the effectiveness score.

\section{Dataset and Performance Evaluation}
To quantitatively demonstrate the effectiveness and efficiency of various image retrieval algorithms, it is indispensable to collect some benchmark datasets and define the evaluation metrics. In this section, we discuss the recent ground truth datasets and distractor datasets used in experimental study for image retrieval. Besides, we introduce the key evaluation indicators in CBIR, including accuracy, efficiency, and memory cost.

\subsection{Recent Dataset for CBIR}
\begin{figure*}
  \centering
  \includegraphics[width = 12 cm]{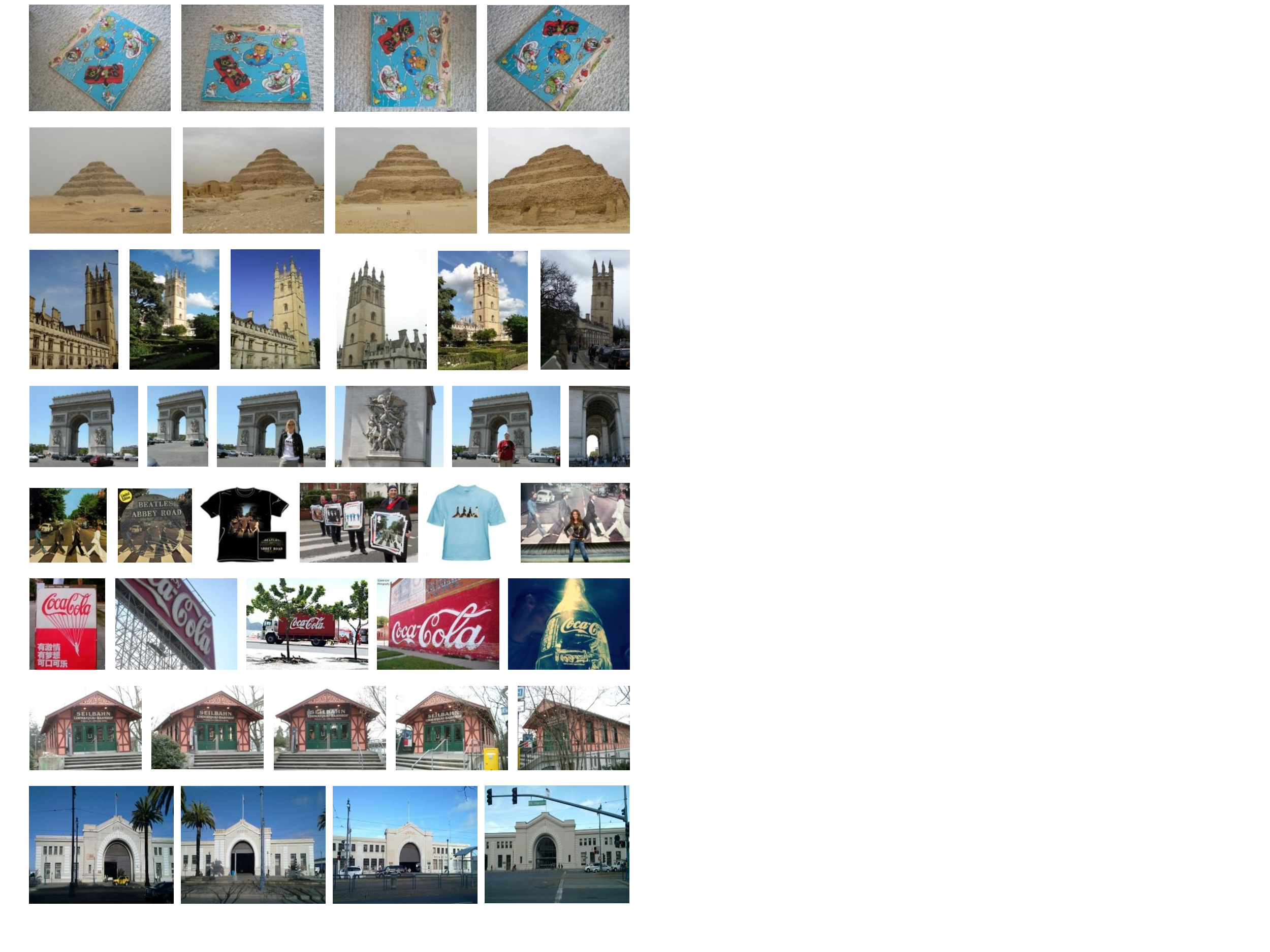}\\
  \caption{Samples images of the existing datasets. First row: UKBench dataset; second row: Holidays dataset; third row: Oxford Building dataset; fourth row: DupImage dataset; fifth row: INSTRE dataset; sixth row: ZuBuD dataset; seventh row: SMVS dataset.}\label{fig:dataSample}
\end{figure*}

Intuitively, the ground-truth dataset should be sufficient large so as to well demonstrate the scalability of image retrieval algorithms. However, considering the tedious labor in dataset collection, the existing ground-truth dataset are relatively small, but mixed with random million-scale distractor database for evaluation on scalability. The existing ground-truth datasets target on particular object/scene retrieval or partial-duplicate Web image retrieval. Generally, the ground-truth images contain a specific object or scene and may undergo various changes and be taken under different views or changes in illumination, scale, rotation, partial occlusion, compression rate,~\emph{etc}. Typical ground truth dataset for this task includes the UKBench dataset~\cite{nister2006scalable}, the Oxford Building dataset~\cite{philbin2007object}, and the Holidays dataset~\cite{jegou2008hamming}, \emph{etc}. MIR Flickr-1M and Flickr-1M are two different million-scale databases which are usually used as distractor to evaluate the scalability of image retrieval algorithms. For convenience of comparison and reference, we list the general information of those recent datasets popularly used in CBIR in Table~\ref{tab:database}. Some sample images from those datasets are shown in Fig.~\ref{fig:dataSample}.

\begin{table*}%
\centering
\caption{General information of the popular retrieval datasets in CBIR. The ``mixed'' database type denotes that the corresponding dataset is a ground truth dataset mixed with distractor images. \label{tab:database}}{%
\begin{tabular}{|l|l|l|l|l|l|}
\hline
Database Name  &Database Type & Database Size & Query Number    & Category Number    & Resolution\\\hline\hline
UKBench        &Ground Truth  & 10,200        & 10,200          & 2,550              & $640 \times 480$ \\\hline
Holidays       &Ground Truth  & 1,491         & 500             & 500                & $1024 \times 768$\\\hline
Oxford-5K      &Mixed         & 6,053         & 55              & 11                 & $1024 \times 768$\\\hline
Paris          &Mixed         & 6,412         & 500             & 12                 & $1024 \times 768$\\\hline
DupImage       &Ground Truth  & 1,104         & 108             & 33                 & $460 \times 350$ (average)\\\hline
FlickrLogos-32 &Mixed         & 8,240         & 500             & 32                 & $1024 \times 768$\\\hline
INSTRE         &Ground Truth  & 28,543        & N/A             & 200                & $1000 \times 720$ (average)\\\hline
ZuBuD          &Ground Truth  & 1,005         & 115             & 200                & $320 \times 240  $\\\hline
SMVS           &Ground Truth  & 1,200         & 3,300           & 1,200              & $640 \times 480$\\\hline
MIR Flickr-1M  &Distractor    & 1,000,000     & N/A             & N/A                & $500 \times 500$ \\\hline
Flickr1M       &Distractor    & 1,000,000     & N/A             & N/A                 & N/A            \\\hline
\end{tabular}}
\end{table*}%

{\bf UKBench dataset} It contains 10,200 images from 2,550 categories\footnote{http://www.vis.uky.edu/$\sim$stewe/ukbench/}. In each category, there are four images taken on the same scene or object from different views or illumination conditions. All the 10,200 images are taken as query and their retrieval performances are averaged.

{\bf Holidays dataset} There are 1,491 images from 500 groups in the Holidays dataset\footnote{http://lear.inrialpes.fr/people/jegou/data.php}. Images in each group are taken on a scene or an object with various viewpoints. The first image in each group is selected as query for evaluation.

{\bf Oxford Building dataset (Oxford-5K)} The Oxford Buildings Dataset\footnote{http://www.robots.ox.ac.uk/$\sim$vgg/data/oxbuildings/} consists of 5062 images collected from Flickr\footnote{http://www.flickr.com/} by searching for particular Oxford landmarks. The collection has been manually annotated to generate a comprehensive ground truth for 11 different landmarks, each represented by 5 possible queries. This gives a set of 55 queries over which an object retrieval system can be evaluated. Some junk images are mixed in it as distractor.

{\bf Paris dataset} In the Paris dataset\footnote{http://www.robots.ox.ac.uk/$\sim$vgg/data/parisbuildings/}, there are 6,412 images, which are collected from Flickr by searching for 12 text queries of particular Paris landmarks. For this dataset, 500 query images are used for evaluation.

{\bf DupImage dataset} This dataset contains 1,104 images from 33 groups\footnote{http://pan.baidu.com/s/1jGETFUm}. Each group corresponds to a logo, a painting, or an artwork, such as KFC, American Gothic Painting, Mona Lisa, \emph{etc}. 108 representative query images are selected from those groups for evaluation.

{\bf FlickrLogos-32 dataset} This dataset\footnote{http://www.multimedia-computing.de/flickrlogos/} contains logo images of 32 different brands which are downloaded from Flickr. All logo images in this dataset have an approximately planar structure. The dataset is partitioned into three subsets for evaluation, \emph{i.e.}, training set, validation set, and query set~\cite{romberg2011scalable}. Of those 8,240 images in the dataset, 6,000 images contain no logos and works as distractors.

{\bf INSTRE} As an instance-level benchmark dataset, the INSTRE dataset~\footnote{http://vipl.ict.ac.cn/isia/instre/} contains two subsets, \emph{i.e.,} INSTRE-S and INSTRE-M~\cite{wang2015instre}. In the former subset, there are 23,070 images, each with a single label of 200 classes. The latter subset contains 5,473 images and each image contains two instances from 100 object categories.


{\bf ZuBuD dataset} The basic dataset contains 1,005 images of 201 buildings in Zurich, with 5 views for each building\footnote{http://www.vision.ee.ethz.ch/showroom/zubud/index.en.html}. Besides, there are additional 115 query images which are not included in the basic dataset. The resolution of those images are uniformly $320 \times 240$.

{\bf Stanford Mobile Visual Search (SMVS) Dataset} This dataset\footnote{http://purl.stanford.edu/rb470rw0983} is targeted for mobile visual search and contains images taken by camera phone on products, CDs, books, outdoor landmarks, business cards, text documents, museum paintings and video clips. It is characterized by rigid objects, widely varying lighting conditions, perspective distortion, foreground and background clutter, and realistic ground-truth reference data~\cite{chandrasekhar2011stanford}. In the dataset, there are 1,200 distinct categories. For each category, one reference image with resolution quality is collected for evaluation. There are 3,300 query images in total which are collected from heterogeneous low and high-end camera phones.

{\bf MIR Flickr-1M} This is a distractor dataset\footnote{http://medialab.liacs.nl/mirflickr/mirflickr1m/}, with one million images randomly downloaded from Flickr and resized to be no larger than 500 by 500.

{\bf Flickr1M} is another distractor database containing SIFT features\footnote{http://bigimbaz.inrialpes.fr/herve/siftgeo1M/} of one million images arbitrarily retrieved from Flickr. The original images in this database are not available.

\subsection{Performance Evaluation for CBIR}
In the design of a multimedia content-based retrieval system, there are three key indicators which should be carefully considered: accuracy, efficiency, and memory cost. Usually, a retrieval method contributes to improving at least one of those indicators with little sacrifice in the other indicators.

{\bf Accuracy}~~To measure the retrieval quality quantitatively, the database images are categorized into difference relevance levels and the accuracy score is summarized based on the rank order of the database images. For different relevance levels, there are different accuracy metrics. Where there are only two relevance level, \emph{i.e.}, relevant and irrelevant, average precision (AP) is widely used to evaluate the retrieval quality of a single query's retrieval results. AP takes consideration of both precision and recall. Precision denotes the fraction of retrieved (top $k$) images that are relevant while recall means fraction of relevant image that are retrieved (in the top $k$ returned results). Generally, for a retrieval system, precision decreases as either the number of images retrieved increases or recall grows. AP averages the precision values from the rank positions where a relevant image was retrieved, as defined in Eq.~\ref{eq:AP}. To summarize the retrieval quality over multiple query images, the mean average precision (mAP) is usually adopted, which average the average precision over all queries.
\begin{equation}
AP = \frac{\sum^n_{k=1}P(k) \cdot rel(k)}{R}
\label{eq:AP}
\end{equation}
where $R$ denotes the number of relevant results  for the current query image, $P(k)$ denotes the precision of top $k$ retrieval results, $rel(k)$ is a binary indicator function equalling 1 when the $k$-th retrieved result is relevant to the current query image and 0 otherwise, and $n$ denotes the total number of retrieved results.

When there are multiple relevance levels, we can resort to normalized discounted cumulative gain (NDCG) metric defined in Eq.~\ref{eq:NDCG} to summarize the ranking results.

\begin{equation}
NDCG = \frac{1}{N}(r_1 + \sum^n_{k=2}\frac{f(r_k)}{log_2(k)}),
\label{eq:NDCG}
\end{equation}
where $n$ denotes the number of retrieved images, $r_k$ denotes the relevance level, $f(\cdot)$ is function to tune the contribution of difference relevance levels, and $N$ denotes the normalized term to ensure that the NDCG score for the ideal retrieved results is 100\%. Popular definitions of $f(\cdot)$ include $f(x) = x$ and $f(x) = 2^x - 1$, with the latter to emphasize on retrieving highly relevant images.

Besides the above measures, some simple measures may be adopted for special dataset. In the public UKBench dataset, considering that there are four relevant images for all queries, the N-S score, \emph{i.e.}, the average 4 times top-4 precision over the dataset, are used to measure the retrieval accuracy~\cite{nister2006scalable}.

{\bf Computational Efficiency}~~The efficiency of a retrieval system involves the time cost in visual vocabulary (codebook) construction, visual feature indexing, and image querying. The first two items are performed off-line, while the last one is conducted on-line. Both the off-line and on-line processing is expected to be as fast as possible. Specially, the on-line querying is usually expected to be responded in real time.

{\bf Memory Cost}~~In a multimedia content-based visual retrieval system, the memory cost usually refers to the memory usage in the on-line query stage. Generally, the memory is mainly spent on the quantizer and the index file of database, which need to be loaded into the main memory for on-line retrieval. Popular quantizer includes tree-based structure, such as hierarchical vocabulary tree, randomized forests, \emph{etc}, which usually cost a few hundred mega-bytes memory for codebook containing million-scale visual words. In some binary code based quantization methods~\cite{zhou2012scalar}~\cite{dong2012high}, the quantizer is simple hash function with negligible memory overhead. For the index file, the memory cost is proportional to the indexed database size. When the database images are represented by local features and each local feature is indexed locally, the index file is proportional to the amount of indexed features and the memory cost per indexed feature.

\section{FUTURE DIRECTIONS}
Despite the extensive research efforts in the past decade, there is still sufficient space to further boost content based visual search. In the following, we will discuss several directions for future research, on which new advance shall be made in the next decade.

\subsection{Ground-Truth Dataset Collection}
In the multimedia and computer vision field, ground-truth datasets are motivated by the specific tasks. At the beginning of those dataset construction, they inspire researchers to update the performance records with their best efforts, leading to many classic ideas and algorithms to address the research problem. However, with the advance to address those datasets, the break-through of some algorithms may suffer from the over-fitting to the dataset. Meanwhile, with deeper understanding and clearer definition of the research problem, the limitation of existing datasets is revealed and new datasets are expected. For content-based image retrieval, we also expect better ground-truth dataset to be collected and released. Generally, the new ground-truth datasets shall be specific to eliminate the ambiguity of relevance of image content, such as logo datasets. Meanwhile, the scale of the dataset shall be sufficiently large so as to distinguish the problem of CBIR from image classification.

\subsection{Intention Oriented Query Formation and Selection}
Intention gap is the first and of the greatest challenge in content-based image retrieval. A simple query in the form of example, color map or sketch map is still insufficient in most time to reflect the user intention, consequently generating unsatisfactory retrieval results. Besides the traditional query formations, assistance from user to specify the concrete expectation will greatly alleviate the difficulty of the following image retrieval process. Considering that the end-users may be reluctant to involve much in the query formation, it is still possible to design convenient query formation interface to reduce the user involvement as much as possible. For instance, it is easy for a user to specify the region of interest in an example image for retrieval, or indicate the expected results are partial-duplicates or just similar in spatial color and texture. It is also possible to predict the potential intentions based on the initial query and make confirmation with end-user. In all, rather than passively induce the intension behind the query, it is beneficial to actively involve end-user in the retrieval process.

In image retrieval, the search performance is significantly impacted by the quality of the query. How to select a suitable query towards the optimal retrieval is a nontrivial issue. The query quality is related with many factors, including resolution, noise pollution, affine distortion, background clutter,~\emph{etc}. In the scenario of mobile search, the query can be selected by guiding the end user to retake better photos. In the server end, automatic retrieval quality assessment methods~\cite{tian2011learning}~\cite{tian2015query} can be designed to select potential candidate from the initial retrieval results of high precision.

\subsection{Deep Learning in CBIR}
Despite the advance in content-based visual retrieval, there is still significant gap towards semantic-aware retrieval from visual content. This is essentially due to the fact that current image representation schemes are hand-crafted and insufficient to capture the semantics. Due to the tremendous diversity and quantity in multimedia visual data, most existing methods are un-supervised. To proceed towards semantic-aware retrieval, scalable  scalable supervised or semi-supervised learning are promising to learn semantic-aware representation so as to boost the content-based retrieval quality. The success of deep learning in large-scale visual recognition~\cite{krizhevsky2012imagenet}~\cite{szegedy2014going}~\cite{simonyan2014very}~\cite{he2014spatial} has already demonstrated such potential.

To adapt those existing deep learning techniques to CBIR, there are several non-trivial issues that deserve research efforts. Firstly, the learned image representation with deep learning shall be flexible and robust to various common changes and transformations, such as rotation and scaling. Since the existing deep learning relies on the convolutional operation with anisotropic filters to convolve images, the resulted feature maps are sensitive to large translation, rotation, and scaling changes. It is still an open problem as whether that can solved by simply including more training samples with diverse transformations. Secondly, since computational efficiency and memory overhead are emphasized in particular in CBIR, it would be beneficial to consider those constraints in the structure design of deep learning networks. For instance, both compact binary semantic hashing codes~\cite{torralba2008small}~\cite{krizhevsky2011using} and very sparse semantic vector representations are desired to represent images, since the latter are efficient in both distance computing and memory storing while the latter is well adapted to the inverted index structure.

\subsection{Unsupervised Database Mining}
In traditional content-based image retrieval algorithms and systems, the database images are processed independently without considering their potential relevance context information. This is primarily due to the fact that, there is usually no label information for the database images and the potential category number is unlimited. Those constraints limit the application of sophisticated supervised learning algorithms in CBIR. However, as long as the database is large, it is likely that there exist some subsets of images and images in each sub-set are relevant to each other images. Therefore, it is feasible to explore the database images with some unsupervised techniques to uncover those sub-sets in the off-line processing stage. If we regard each database image as a node and the relevance level between images as edge to link images, the whole image database can be represented as a large graph. Then, the sub-sets mining problem can be formulated as a sub-graph discovery problem. On the other hand, in practice, new images may be incrementally included into the graph, which casts challenge to dynamically uncover those sub-graphs on the fly. The mining results in the off-line stage will be beneficial for the on-line query to yield better retrieval results.

\subsection{Cross-modal Retrieval}
In the above discussion of this survey, we focus on the visual content for image retrieval. However, besides the visual features, there are other very useful clues, such as the textual information around images in Web pages, the click log of users when using the search engines, the speech information in videos,~\emph{etc}. Those multi-modal clues are complementary to each to collaboratively identify the visual content of images and videos. Therefore, it would be beneficial to explore cross-modal retrieval and fuse those multi-modal features with different models. With multi-modal representation, there are still many open search topics in terms of collaborative quantization, indexing, search re-ranking, \emph{etc}.

\subsection{End-to-End Retrieval Framework}
As discussed in the above sections, the retrieval framework is involved with multiple modules, including feature extraction, codebook learning, feature quantization, feature quantization, image indexing, \emph{etc}. Those modules are individually designed and independently optimized for the retrieval task. On the other hand, if we investigate the structure of the convolutional neural network (CNN) in deep learning, we can find a very close analogy between the BoW model and the CNN model. The convolutional filters used in the CNN model works in a similar way as the codewords of the codebook in the BoW model. The convolution results between the image patch and the convolution filter are essentially the soft quantization results, with the max-pooling operation similar to the local aggregation in the BoW model. As long as the learned feature vector is sparse, we can also adopt the inverted index structure to efficiently index the image database. Different from the BoW model, the above modules in the CNN model are collaboratively optimized for the task of image classification. Based on the above analogy, similarly, we may also resort to an end-to-end paradigm to design a framework that takes images as input and outputs the index-oriented features directly, with the traditional key retrieval-related modules implicitly and collaboratively optimized.

\subsection{Social Media Mining with CBIR}
Different from the traditional unstructured Web media, the emerging social media in recent years have been characterized by community based personalized content creation, sharing, and interaction. There are many successful prominent platforms of social media, such as Facebook, Twitter, Wikipedia, LinkedIn, Pinterest, \emph{etc}. The social media is enriched with tremendous information which dynamically reflects the social and cultural background and trend of the community. Besides, it also reveals the personal affection and behavior characteristics. As an important media of the user-created content, the visual data can be used as an entry point with the content-based image retrieval technique to uncover and understand the underlying community structure. It would be beneficial to understand the behavior of individual users and conduct recommendation of products and services to users. Moreover, it is feasible to analyze the sentiment of crowd for supervision and forewarning.

\subsection{Open Grand Challenge}
Due to the difference in deployment structure and availability of data, the research on content based image retrieval in the academia suffers a gap from the real application in industry. To bridge this gap, it is beneficial to initiate some open grand challenges from the industry and involve the researchers in the academia to investigate the key difficulties in real scenarios. In the past five years, there are some limited open grand challenge, such as the Microsoft Image Grand Challenge on Image Retrieval\footnote{http://acmmm13.org/submissions/call-for-multimedia-grand-challenge-solutions/msr-bing-grand-challenge-on-image-retrieval-scientific-track} and Alibaba Large-Scale Image Search Challenge\footnote{http://tianchi.aliyun.com/competition/introduction.htm?spm=5176.100069\\.5678.1.SmufkG$\&$raceId=231510$\&\_$lang=en$\_$US}. In the future, we would expect many more such grand challenges. The open grand challenge will only only advance the research progress in the academia, but also benefit the industry with more and better practical and feasible solutions to the real-world challenges.

\section{CONCLUSIONS}
In this paper, we have investigated the advance on content-based image retrieval in recent years. We focus on the five key modules of the general framework,~\emph{i.e.}, query formation, image representation, image indexing, retrieval scoring, and search re-ranking. For each component, we have discussed the key problems and categorized a variety of representative strategies and methods. Further, we have summarized eight potential directions that may boost the advance of content based image retrieval in the near future.


\small
\bibliographystyle{IEEEtran}
\bibliography{cbir_ref}

%




\end{document}